\documentclass[letter]{aa} 

\usepackage{graphicx}
\usepackage{txfonts}
\usepackage{multirow}
\usepackage{colortbl}

\usepackage{hyperref}  
\hypersetup{colorlinks=true,linkcolor=[rgb]{1.,0.2,0.2},citecolor=[rgb]{0.1,0.4,1.},filecolor=[rgb]{0.7,0.2,0.2},urlcolor=[rgb]{0.7,0.2,0.2}}
\usepackage{color}
\definecolor{blue}{rgb}{0., 0., 1}

\newcommand{\hii}{\textrm{H}\textsc{ii}}

\newcommand{\ha}{\ifmmode {\rm H}\alpha \else H$\alpha$\fi}
\newcommand{\halam}{\ifmmode {\rm H}\alpha \lambda6563 \else H$\alpha$ $\lambda$6563 \fi}
\newcommand{\hb}{\ifmmode {\rm H}\beta \else H$\beta$\fi}
\newcommand{\hg}{\ifmmode {\rm H}\gamma \else H$\gamma$\fi}
\newcommand{\hblam}{\ifmmode {\rm H}\beta \lambda4861 \else H$\beta$ $\lambda$4861 \fi}
\newcommand{\lya}{\ifmmode {\rm Ly}\alpha \else Ly$\alpha$\fi}
\newcommand{\pg}{\ifmmode {\rm P}\gamma \else Pa$\gamma$\fi}
\newcommand{\lyb}{\ifmmode {\rm Ly}\beta \else Ly$\beta$\fi}
\newcommand{\lyg}{\ifmmode {\rm Ly}\gamma \else Ly$\gamma$\fi}

\newcommand{\flyc}{\ifmmode  \mathrm{f}_\mathrm{esc}\mathrm{(LyC)} \else $\mathrm{f}_\mathrm{esc}\mathrm{(LyC)}$\fi}

\def\kms{km s$^{-1}$}

\def\ergs{\ifmmode \mathrm{erg\hspace{1mm}s}^{-1} \else erg s$^{-1}$\fi}

\def\micron{\ifmmode \mu\mathrm{m} \else $\mu$m\fi}
\def\msun{\ifmmode \mathrm{M}_{\odot} \else M$_{\odot}$\fi}
\def\msunyr{\ifmmode \mathrm{M}_{\odot} \hspace{1mm}{\rm yr}^{-1} \else $\mathrm{M}_{\odot}$ yr$^{-1}$\fi}
\def\zsun{\ifmmode Z_{\odot} \else Z$_{\odot}$\fi}
\def\lsun{\ifmmode L_{\odot} \else L$_{\odot}$\fi}
\def\mstar{\ifmmode \mathrm{M}_{\star} \else M$_{\star}$\fi}
\newcommand{\JWST}{\textrm{JWST}}

\newcommand\TM{M25}
\defcitealias{Morishita2025}{\TM}

\usepackage{orcidlink}

\newcommand{\orcid}[1]{\href{https://orcid.org/#1}{\textcolor[HTML]{A6CE39}{\aiOrcid}}}

\begin{document}

\titlerunning{An extremely faint and low-metallicity Lyman-continuum leaker candidate in the Epoch of Reionization}

\title{A faint  M$_{\rm UV} = -14.5$ Lyman-continuum leaker candidate in the epoch of reionization: Unprecedented Ly$\alpha$ properties at z=5.725}

\authorrunning{Matteo Messa et al.}
\author{
M.~Messa\inst{\ref{inafbo}}\fnmsep\thanks{E-mail: \href{mailto:matteo.messa@inaf.it}{matteo.messa@inaf.it}}$^{\orcidlink{0000-0003-1427-2456}}$ \and
E.~Vanzella\inst{\ref{inafbo}}$^{\orcidlink{0000-0002-5057-135X}}$ \and
T.~Morishita\inst{\ref{ipac}, \ref{tohoku}}$^{\orcidlink{0000-0002-8512-1404}}$ \and
M.~Stiavelli\inst{\ref{stsci}}$^{\orcidlink{0000-0001-9935-6047}}$ \and
T.~Treu \inst{\ref{UCLA}}$^{\orcidlink{0000-0002-8460-0390}}$ \and
P.~Bergamini\inst{\ref{inafbo}}$^{\orcidlink{0000-0003-1383-9414}}$\and
Z.~Liu\inst{\ref{mitkavli}, \ref{gradsc_tohoku}}$^{\orcidlink{0009-0002-8965-1303}}$\and
A.~Zanella\inst{\ref{inafbo}}$^{\orcidlink{0000-0001-8600-7008}}$\and
A.~Bolamperti\inst{\ref{mpa}}$^{\orcidlink{0000-0001-5976-9728}}$\and
A.~Verhamme\inst{\ref{CRAL},\ref{obsgeneve}}$^{\orcidlink{0000-0002-2201-1865}}$\and
T.~Garel\inst{\ref{obsgeneve},\ref{CRAL}}$^{\orcidlink{0000-0002-9613-9044}}$
C.~Grillo\inst{\ref{unimi},\ref{inafiasf}}$^{\orcidlink{0000-0002-5926-7143}}$\and
P.~Rosati \inst{\ref{unife},\ref{inafbo}}$^{\orcidlink{0000-0002-6813-0632}}$
}
\institute{
INAF -- OAS, Osservatorio di Astrofisica e Scienza dello Spazio di Bologna, via Gobetti 93/3, I-40129 Bologna, Italy \label{inafbo} 
\and
IPAC, California Institute of Technology, Pasadena, CA 91125, USA \label{ipac} 
\and
Astronomical Institute, Tohoku University, 6-3 Aramaki, Aoba-ku, Sendai 980-8578, Japan\label{tohoku}
\and 
Space Telescope Science Institute (STScI), 3700 San Martin Drive, Baltimore, MD 21218, USA\label{stsci}
\and
Department of Physics and Astronomy, University of California, Los Angeles, 430 Portola Plaza, Los Angeles, CA 90095, USA\label{UCLA}
\and
MIT Kavli Institute for Astrophysics and Space Research, 70 Vassar Street, Cambridge, MA 02139, USA\label{mitkavli}
\and
Astronomical Institute, Graduate School of Science, Tohoku University, Sendai, Miyagi 980-8578, Japan\label{gradsc_tohoku}
\and
Max-Planck-Institut f\"ur Astrophysik, Karl-Schwarzschild-Str. 1, D-85748 Garching, Germany\label{mpa}
\and
Univ Lyon, Univ Lyon1, ENS de Lyon, CNRS, Centre de Recherche Astrophysique de Lyon UMR5574, Saint-Genis-Laval, France\label{CRAL}
\and
Observatoire de Genève, Université de Genève, Chemin Pegasi 51, 1290 Versoix, Switzerland\label{obsgeneve}
\and
Dipartimento di Fisica, Università degli Studi di Milano, Via Celoria 16, I-20133 Milano, Italy\label{unimi}
\and 
INAF -- IASF Milano, via A. Corti 12, I-20133 Milano, Italy\label{inafiasf}
\and
Dipartimento di Fisica e Scienze della Terra, Università degli Studi di Ferrara, Via Saragat 1, I-44122 Ferrara, Italy\label{unife}
}

\date{} 
 
\abstract
{
We report the unprecedented Ly$\alpha$ properties of AMORE6, an extremely metal-poor ($12+\log({\rm O/H}) < 6$), low-mass ($M_\star = 4.4\times10^{5}\,M_\odot$), and ultracompact (effective radius $\sim30$ pc) dwarf galaxy at $z=5.7253$, which is gravitationally lensed by the cluster A2744. A prominent, narrow, and nearly symmetric Ly$\alpha$ emission line is detected at the systemic redshift (the latter traced by H$\beta$, from \JWST/NIRCam slitless spectroscopy), with a rest-frame equivalent width of $150 \pm 10$ \AA   , a full width at half maximum of  $\rm 58\pm1$ km~s$^{-1}$, and a slight asymmetry, resulting in a flux excess of $\rm \sim10\%$ in the red wing of the line. The negligible velocity offset from systemic ($dv = 4\pm67$ km s$^{-1}$, $3\sigma$ uncertainty), together with the sharpness and symmetry of the profile, indicates minimum radiative transfer effects, which implies a neutral hydrogen column density consistent with an optically thin medium that in turn is compatible with a nonzero ionizing photon escape fraction. If indirect spectral diagnostics calibrated at $z<4.5$ remain the only viable tools for identifying LyC leakers during reionization, then based on its strongest indicator (Ly$\alpha$), AMORE6 stands out as one of the most compelling LyC-leaking candidates yet discovered in the epoch of reionization.
}
\keywords{galaxies: high-redshift -- galaxies: star formation -- stars: Population III -- gravitational lensing: strong.}
   \maketitle

\section{Introduction}
\label{sect:intro}

The epoch of reionization (EoR) marks the last main phase transition of the Universe, when the first stars and galaxies revealed their presence by ionizing the surrounding neutral hydrogen (H\,{\sc i}; e.g., \citealt{Loeb&Furlanetto2013}). While the reionization timeline is increasingly well constrained ($z \simeq 6$--9; e.g., \citealp{Fan2006,Mason2019,Mason2026,PlanckVI_2020}), the nature of the dominant ionizing sources remains elusive.
Faint, low-mass galaxies have been proposed as the primary contributors to the ionizing photon budget \citep[e.g.,][]{Finkelstein2019,Simmonds2024}, although the role of massive galaxies is not yet fully understood, as they may also produce substantial Lyman continuum (LyC) emission \citep[e.g.,][]{naidu20,MarquesChaves2024}. Direct measurements of ionizing UV radiation ($\lambda < 912$\AA) are prevented at $z \gtrsim 4.5$ by the high opacity of the intergalactic medium \citep[IGM][]{worseck14,Inoue2014}. Thus, current efforts focus on low-redshift LyC emitters \citep[e.g.,][]{Flury2022a,Jaskot2024a} as analogs of $z>6$ galaxies to calibrate indirect diagnostics of LyC leakage applicable in the EoR.
Recent work highlighted the multiparameter nature of LyC leakage, requiring correlations with diverse physical properties to be established through multivariate approaches. For instance, \citet{Jaskot2024a}, analyzed 35 local ($z\sim0.3$) LyC emitters \citep{Flury2022a,Flury2022b} and identified several key diagnostics: the EW of Lyman absorption features, the UV $\beta$ slope \citep[see also][]{Chisholm2022}, \lya\ peak separation and shift, the Ly$\alpha$ escape fraction \citep[see also][]{Dijkstra2016}, dust excess $E(B-V)_{\rm neb}$, the star formation rate surface density, and the O32 index ([OIII]4959,5007 / [OII]3727). These diagnostics have also been applied to a handful of confirmed LyC emitters at intermediate redshift ($1<z<4$; \citealp{Jaskot2024b}),
and they now provide a framework for probing reionization-era sources, particularly with JWST access to rest-frame optical lines at $z=6$--9 \citep[e.g.,][]{Mascia2024a}. 
In parallel, radiative transfer models predict a tight connection between LyC escape and Ly$\alpha$ spectral properties \citep[e.g.,][]{behrens14,Verhamme2015,Monter2026}. Observations and radiative transfer models confirm that galaxies with multipeaked Ly$\alpha$ profiles and a narrow peak close to systemic are likely optically thin to LyC \citep[e.g.,][]{schaerer11grid,Verhamme2017,Verhamme2018,Izotov2018,izotov21lowmass,Vanzella2020_ion2,Naidu2022}. The Ly$\alpha$ peak separation is controlled by the residual H\,{\sc i}
column density of the carved channels, and the line asymmetry correlates with the porosity and multiphase structure of the
\hii\ region \citep[e.g.,][]{Kakiichi2021}.  
However, the increasing IGM neutrality during the EoR (partially) suppresses and reshapes Ly$\alpha$, limiting its diagnostic power \citep[e.g.,][]{Pentericci2014,Garel2021,Mason2026}. Mitigating cases arise for sources located in ionized bubbles around luminous $z\simeq6$ quasars \citep[e.g.,][]{Protusova2025} or along unusually transparent sightlines \citep{Matthee2018}, where double-peaked Ly$\alpha$ emission with a narrow peak separation has been observed up to $z\simeq 6.5$, pointing to nonzero LyC escape. 

In this Letter, we revisit the extremely metal-poor source AMORE6, which was recently identified by \citet{Morishita2025} with $12+\log({\rm O/H}) < 6$, in the context of escaping ionizing radiation. AMORE6 is a remarkably faint ($M_{1700} \simeq -14.5$), compact ($\rm R_{eff}\sim30~pc$) galaxy at $z=5.725$ that is strongly magnified by the galaxy cluster A2744. Beyond its extraordinary low metallicity and luminosity, its unprecedented Ly$\alpha$ spectral properties  provide compelling evidence for substantial ionizing photon escape at $z\simeq 6$,  opening a new window on the nature of reionization-era galaxies. Throughout the Letter, we assume a flat cosmology with $\Omega_{\rm M} = 0.3$, $\Omega_{\rm \Lambda} = 0.7$, and $H_{\rm 0} = 70\,{\rm km}\,{\rm s}^{-1}\,{\rm Mpc}^{-1}$.  All magnitudes are given in the AB system \citep{Oke_1983}: $m_{\rm AB} = 23.9 - 2.5 \log(f_\nu / \mu{\rm Jy})$.

\begin{figure*}
\center
 \includegraphics[width=0.85\textwidth]{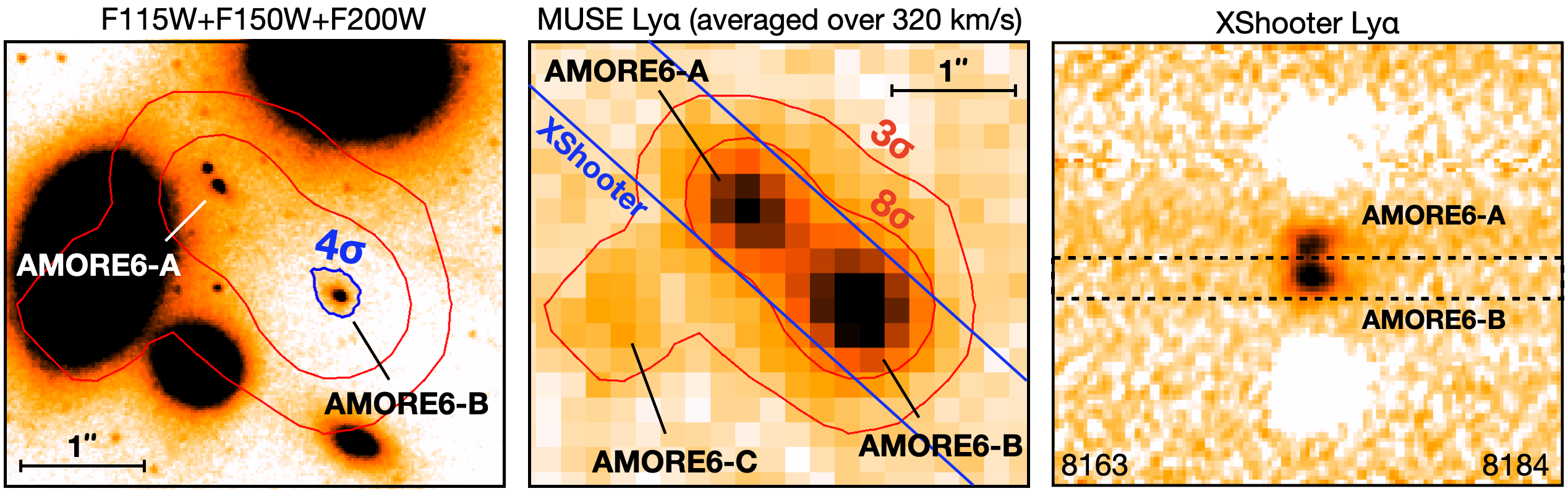}
 \caption{Left: NIRCam stacking of the SW bands. The red contours highlight the extent of the \lya\ emission, and the blue contour encloses the rest-UV emission of AMORE6-B. Middle: \lya\ flux from VLT/MUSE; a faint emission from a low-magnification third counter-image (AMORE6-C) is also visible. Right: Zoomed two-dimensional X-shooter spectrum centered on the \lya\ wavelength from 8163~\AA\ to 8184~\AA. The dotted lines show the region we used to extract spectra.}
 \label{fig:imaging}
\end{figure*}
\begin{figure}
\center
 \includegraphics[width=\columnwidth]{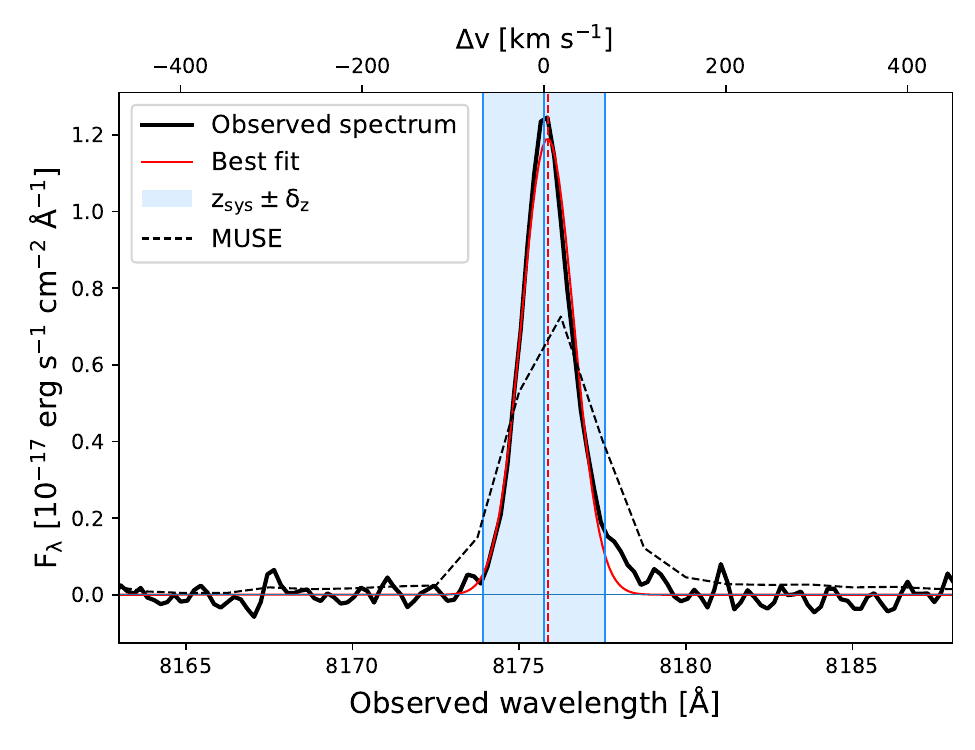}
 \caption{Observed profile of the \lya\ emission for AMORE6-B (black line), best-fit profile (solid~red), best-fit central wavelength (dashed red), systemic redshift (and relative uncertainty) from \citet[solid blue lines and shaded region]{Morishita2025}. The (lower-resolution) MUSE spectrum is shown as the dashed black line.
 } 
 \label{fig:line}
\end{figure}

\section{Analysis} \label{sec:analysis}
The analysis of the rest-optical spectrum\footnote{from the NIRCam WFSS Cycle 2 program “All the Little Things", GO 3516; PIs: Matthee \& Naidu, \citep[see][]{Naidu2024_arXiv}}, in particular, the H$\beta$ emission, was presented by \citealt{Morishita2025} (hereafter \citetalias{Morishita2025}). The H$\beta$ redshift from \citetalias{Morishita2025} is adopted here as the systemic redshift of AMORE6 (see Table~\ref{tab:properties}); in a conservative approach, we considered its $3\sigma$ uncertainty. 
From the same analysis, the UV magnitude of the AMORE6-B counter-image and its UV $\beta$ slope were used to estimate the expected continuum at the \lya\ wavelength. The flux from image A is contaminated by the light from three foreground sources; for this reason, its magnitude was not reported by \citetalias{Morishita2025}. In addition, image B has the strongest magnification ($\mu \simeq 77$, compared to $\mu \simeq 39$ for AMORE6-A; \citealt{Bergamini2023}), and thus, we considered AMORE6-B as the reference for this spectral analysis; we verified that the main results of this work remain unchanged when A is considered instead (see Table~\ref{tab:properties}).

The analysis of the rest-UV spectrum was based on observations from the Multi Unit Spectroscopic Explorer (MUSE) and X-shooter instruments, on the Very Large Telescope (VLT). The MUSE data we used were described by \citet{mahler18} and \citet{Bergamini2023}. 
X-shooter observations (PI T. Morishita, DDT, prog.id. 115.29G6.001) were carried out between August and September 2025 with a median seeing of $1.0_{-0.2}^{+0.1}$ arcsec for a total integration time on source of 11 h. The full dataset will be presented in a forthcoming work (Morishita et al. in prep.). Briefly, we used science ESO level 2 products, specifically focusing on the visual arm (VIS), where the \lya\ lies. The \lya\ emission is clearly detected in each OB with a signal-to-noise ratio S/N~$>15$. 
All OBs were visually inspected and aligned in four cases along the spatial direction by 1 pixel (measuring the \lya\ peak), 
averaged,
and the one-dimensional spectrum was extracted by centering on the source AMORE6-B from a window enclosing 10 spatial pixels ($1.6''$; Fig.~\ref{fig:imaging}). The spectrum was converted from air into vacuum reference wavelength and was then corrected to the barycentric radial velocity\footnote{To make it consistent with the NIRCam-WFSS spectrum of \citetalias{Morishita2025}.} (through {\tt astropy}). The resulting \lya\ emission sampled at a spectral resolution R~$\simeq8900$ is shown in Fig.~\ref{fig:line}. The line location is fully consistent with the location inferred from MUSE (see Table~\ref{tab:properties}). The MUSE spectrum was also used to calibrate the X-shooter \lya\ flux, in order to correct for the slit loss. 

We fit the line with {\tt specutils}/{\tt astropy} using Gaussian profiles. The fit quantities are reported in Table~\ref{tab:properties}. 

\begin{table}[]
    \centering
    \caption{\lya-related properties of AMORE6.} 
    \renewcommand{\arraystretch}{1.35}
    \begin{tabular}{l|ccc}
    & B & A & B (MUSE) \\
    \hline
    $\rm z_{sys}^{(a)}$              & $5.7253$                  & $5.7253$                  & $5.7253$ \\
    $\rm \delta(z_{sys})^{(a)}$      & $0.0015$                  & $0.0030$                  & $0.0015$ \\
    $\rm z_{\lya}$                   & $5.7254$                  & $5.7256$                  & $5.7256$ \\
    $\rm \delta(z_{\lya})^{(b)}$     & $0.0001$                  & $0.0001$                  & $0.0001$ \\
    $d v$ $\rm [km~s^{-1}]$          & $4\pm67$                  & $13\pm147$                & $13\pm147$ \\
    $\rm FWHM_{\lya}~[\AA]^{(c)}$    & $1.83\pm0.03$             & $1.81\pm0.04$             & $3.06\pm0.05$ \\
    $\rm FWHM_{\lya}~\left[\frac{km}{s}\right]^{(d)}$   & $58.2\pm1.4$      & $57.1\pm1.7$              & $63.6\pm3.2$ \\
    $\rm A_{RB}$$\rm ^{(e)}$               & $0.12^{+0.04}_{-0.05}$ & $0.19^{+0.05}_{-0.08}$ & $0.31^{+0.01}_{-0.01}$ \\
    $\rm A_{PERC}$$\rm ^{(e)}$             & $0.11^{+0.06}_{-0.06}$ & $0.24^{+0.10}_{-0.10}$ & $0.06^{+0.01}_{-0.01}$ \\
    $\rm f_{\lya}~[10^{-17}cgs]$     & $2.32\pm0.05$             & $1.72\pm0.06$             & $2.32\pm0.05$ \\
    $\rm mag_{\lya,continuum}$$\rm ^{(f)}$ & $27.14^{+0.07}_{-0.08}$ & $-$                   & $27.14^{+0.07}_{-0.08}$ \\
    $\rm EW_{\lya}~[\AA]$            & $150^{+10}_{-12}$         & $-$                       & $150^{+10}_{-12}$ \\
    \hline
    \end{tabular}
    \tablefoot{The properties in Cols. 1 and 2 are taken from X-shooter data for images B and A, respectively; MUSE measurements for image B are also reported in Col. 3. (a) From \citetalias{Morishita2025}; the uncertainty here is 
    $3\sigma$. NIRCam-WFSS at observed \hb\ has $\rm R\sim1500$, implying  $\sigma_z \sim 0.004$, or $\rm 
    \sigma_v\sim 200~km s^{-1}$. (b) $5\sigma$ uncertainty (see text). (c) Observed FWHM. (d) Instrument-
    corrected FWHM. (e) RB and PERC asymmetry as defined in the text. (f) $\rm M_{UV}$ from 
    \citetalias{Morishita2025} extrapolated to the observed \lya\ continuum using $\beta = -2.77$.}
    \label{tab:properties}
\end{table}


\section{Results} \label{sec:results}
Figure~\ref{fig:line} shows the VLT/X-shooter Ly$\alpha$ line extracted from AMORE6-B; the line is detected at an S/N~$\simeq40$, centered at $\lambda = 8175.81 \pm 0.08$~\AA\ (vacuum), corresponding to $z = 5.7254\pm0.0001$ (for a rest-frame Ly$\alpha$ wavelength of 1215.67~\AA). The rest-frame equivalent width is $150 \pm 10$~\AA. Two unprecedented properties emerge at this high redshift: (1) the velocity offset with respect the systemic redshift is $dv = 4\pm 67$ \kms, which is fully consistent with emission located at the resonance frequency ($dv=0$); (2) even at the high spectral resolution of X-shooter ($d\lambda\simeq0.92$~\AA, corresponding to $\rm \simeq34~km s^{-1}$), the line profile is nearly symmetric and narrow, with a measured FWHM of $1.83\pm0.03$~\AA, or $58\pm1$ \kms, after correcting for instrumental broadening. The line is slightly asymmetric, which we quantified with the red/blue asymmetry parameter ($\rm A_{RB}$), that is, the flux difference redward and blueward of the observed line peak, normalized by the line flux; the value $\rm A_{RB}=0.12^{+0.04}_{-0.05}$ indicates a flux excess of $\sim10\%$ on the red side of the line. 
Similar results emerge from the nonparametric percentile asymmetry parameter, defined as A$_{\rm PERC} = \frac{(\lambda_{90} - \lambda_{50})-(\lambda_{50}-\lambda_{10})}{(\lambda_{90}-\lambda_{10})}$, which shows A$_{\rm PERC} = 0.11\pm0.06$, which is still close to symmetry with a faint excess toward the red side.
We point out that overall consistent results were obtained when the \lya\ from the AMORE6-A image was considered (Table~\ref{tab:properties} for details). Although the spectral resolution is lower by 2.5 times, the analysis of the MUSE spectrum also recovers similar properties (Table~\ref{tab:properties}).

These properties have never been observed so clearly at this redshift.
Recently, \citet{Saxena2024} and \citet{PrietoLyon2025_arXiv} identified few galaxies showing an emerging \lya\ line at the systemic redshift (i.e., very small $dv$), within a large sample of Lyman-alpha emitters (LAEs) at $5<z<8$ (Appendix~\ref{sec:app:lensed_sample}). However, in most cases, the line profile is larger than what is measured in AMORE6 ($\rm FWHM\gtrsim200~km~s^{-1}$) and the asymmetry is very pronounced, indicating the effect of gas reprocessing of the line. In the remaining cases, the S/N is too low for us to infer the line profiles. We point out that $z\gtrsim4$ LAEs studied in the literature (including other galaxies magnified by gravitational lensing) are brighter by about 4-6 magnitudes than AMORE6, M$_{\rm UV} \lesssim -18$ (see Appendix~\ref{sec:app:lensed_sample}).
The comparison with confirmed LyC leakers at lower redshifts ($\rm z\leq4$) showing Ly$\alpha$ at the systemic velocity (e.g., Sunburst, \citealt{rivera19} and \citealt{Vanzella2022_sunburst}; Ion3, \citealt{Vanzella2018_ion3} and \citealt{Mestric2025}; Ion2, \citealt{Vanzella2020_ion2}), together with radiative transfer models (see below) indicates that AMORE6 might be the most promising LyC-leaking candidate known to date at $z\simeq6$.

In dense H\,{\sc i} environments, \lya\ photons scatter until they shift out of resonance, producing broad profiles with little flux at systemic and widely separated peaks. In contrast, in AMORE6, all of the \lya\ flux emerges close to resonance, suggesting extremely low column densities and likely ionized channels that perforate the ISM. 
In particular, given the steep ultraviolet slope ($\beta = -2.77$, F$_{\lambda} \propto \lambda^{\beta}$) and the extremely low metallicity ($12+\log({\rm O/H}) < 6$; \citetalias{Morishita2025}), the observed \lya\ emission is expected to be negligibly affected by dust attenuation and only weakly affected by neutral gas damping. A similar conclusion can be inferred from radiative transfer (RT) models \citep[e.g.,][]{behrens14,AlmadaMonter2024}, which show that densities as low as $N_{\rm HI} \gtrsim 10^{16}$ cm$^{-2}$ are reflected in asymmetric and/or multipeaked \lya\ profiles, displaced from the systemic redshift (in the expanding-shell scenario, with low to moderate expansion velocities, $\rm v_{exp}\lesssim100~km s^{-1}$; \citealp{Verhamme2015,Verhamme2018}; see our  Appendix~\ref{sec:app:RT}).
The optical depth of \lya\ photons is indeed higher by $\sim 10^4$ times than that of LyC photons \citep[e.g.,][]{Verhamme2015}, and the detection of copious \lya\ emission that peaks at the line center would therefore indicate column densities that are sufficiently low to imply a large escape fraction of ionizing photons ($f_{\rm esc, LyC} \simeq 0.5$–1.0). 

The low column density we derived might imply an ionized channel along the line of sight. Alternatively, the observed line profile might be the sign of a clumpy ISM with a low covering fraction. A perforated dense neutral ISM may still give rise to a double-peak profile that surrounds the central profle \citep[e.g.,][]{rivera17,AlmadaMonter2024,Monter2026}, as observed in some confirmed leakers, \citep[e.g.,][]{rivera19,Vanzella2018_ion3,Vanzella2020_ion2}. The absence of multiple \lya\ peaks in the observed spectrum suggests that the ISM around AMORE6 has a low covering fraction overall. 
From simulations that coupled \lya\ transfer with detailed radiation hydrodynamics in individual \hii\ regions, \citet{Kakiichi2021} studied the interplay between \lya\ and LyC escape and suggested that the shape of the main \lya\ peak can distinguish between anisotropic LyC leakage through holes in a turbulent \hii\ region (indicated by asymmetry within the red peak in a double-peaked profile) and isotropic LyC leakage from a fully density-bounded \hii\ region (symmetric profile).
AMORE6 would then represent the first likely example of an isotropic LyC emitter at $z\simeq 6$.

In this likely scenario, the narrow \lya\ line resolved with X-shooter and its nearly symmetric profile suggest that the blue wing is only weakly affected by absorption of the intergalactic medium, which at $z=5.7$ might otherwise attenuate more than 50\% of the total line flux \citep[e.g.,][]{Laursen2011}. As reported by \citetalias{Morishita2025}, AMORE6 lies between two known overdensities that are each separated by $\sim 5$ proper Mpc from the source. These structures might have generated an ionized region around AMORE6 that mitigates the expected IGM damping.

As a alternative to the main interpretation, RT models predict a relatively symmetric emission at systemic redshift at higher densities ($\rm N_{\rm HI} > 10^{17}$ cm$^{-2}$) in case of fast-expanding gas shells ($\rm v_{exp}\gg100~km\,s^{-1}$; \citealp{schaerer11grid}; see our Appendix~\ref{sec:app:RT}). Fast outflows like this are observed in massive and/or highly star-forming galaxies ($\rm M\gtrsim10^9~M_\odot$, $\rm SFR\gtrsim 1~M_\odot yr^{-1}$; \citealp{Chisholm2015}). However, we recall that AMORE6 is intrinsically a small and low-mass system, that is, it is closer to a massive star cluster or compact HII region, for which velocities $\rm <100~km\,s^{-1}$ are expected \citep[e.g.,][]{TenorioTagle2015,Turner2015}. 
In addition, studies of local LyC emitters found that LAE systems with the narrowest \lya\ profile (as in the case of AMORE6) show the lowest velocities \citep[e.g.,][]{Jaskot2017}.

Finally, a further possibility to explain the observed \lya\ emission at $dv\approx0~\rm{km~s^{-1}}$ in case of high densities is a shift in the actual systemic redshift of AMORE6 by $\rm \sim100-200~km\,s^{-1}$; such a substantial difference would still be consistent with the redshift determined by \citetalias{Morishita2025} within $3\sigma$, given the uncertainty associated with the \hb\ line center (see Table~\ref{tab:properties}). 
In this case, the observed \lya\ would be a gas-processed red peak, where the absence of a blue peak might be attributed to the suppression by absorption from the IGM \citep[e.g.][]{Inoue2014,Garel2021}. 
We stress, however, that even in this case, for gas densities $\rm >10^{17}~cm^{-2}$, the observed line would broaden and likely become asymmetric (as shown in Appendix~\ref{sec:app:RT}); the narrow and almost symmetric profile of the \lya\ line that is robustly characterized at high spectral resolution strongly disfavor this scenario.

\section{Final remarks} \label{sec:remarks}
AMORE6 is an exceptional source for several reasons: (1) its currently inferred metallicity is the lowest known among high-redshift galaxies, making it a prime candidate pristine star-forming region \citepalias[][]{Morishita2025}; (2) it is strongly lensed by a factor of $\sim77$ \citep[][]{Bergamini2023} and can therefore be spatially resolved down to $\rm \sim30~pc$ in the image plane; and (3) it represents the most compelling case of a Lyman-continuum leaker candidate identified during the epoch of reionization (along with the $\rm z\sim10$ candidate from \citealp{Marques-Chaves2026}). Its \lya\ emission properties are consistent with a HI column optically thin to LyC, which is indicative of a high global escape fraction.

Under these conditions, in which little (or no) radiative transfer reprocesses the \lya\ line, its profile is expected to be consistent with the Balmer lines emitted from the same system. Unfortunately, as discussed in \citetalias{Morishita2025}, the current low spectral resolution of \hb\ from NIRCam-WFSS prevents a detailed comparison between the two.
Deep NIRSpec observations with a high spectral resolution are needed to characterize the shape of the Balmer lines (and possibly measure the gas expansion velocity), to refine the systemic redshift, and to search for additional emission lines.

\begin{acknowledgements}
We thank the anonymous referee for the useful comments and suggestions which helped improving the draft. 
MM and EV acknowledge financial support through grants INAF GO Grant 2022 ``The revolution is around the corner: JWST will probe globular cluster precursors and Population III stellar clusters at cosmic dawn'' and INAF GO Grant 2024 ``Mapping Star Cluster Feedback in a Galaxy 450 Myr after the Big Bang'', and by the European Union – NextGenerationEU within PRIN 2022 project n.20229YBSAN - Globular clusters in cosmological simulations and lensed fields: from their birth to the present epoch. TM received support from NASA through the STScI grant JWST-GO-3990. MS acknowledges partial support through NASA grant 80NSSC21K1294.
This research has used NASA’s Astrophysics Data System, QFitsView, and SAOImageDS9, developed by Smithsonian Astrophysical Observatory.
Additionally, this work made use of the following open-source packages for Python, and we are thankful to the developers of these: Matplotlib \citep{matplotlib2007}, 
Numpy \citep[][]{NUMPY2011}, 
Astropy \citep{astropy:2022} (http://www.astropy.org).
\end{acknowledgements}

%
%

\bibliographystyle{aa}
\bibliography{bib}

\begin{appendix} 
\onecolumn
\section{Literature samples of LAEs at z>4.}\label{sec:app:lensed_sample}
\begin{figure*}[h!]
\includegraphics[width=0.33\textwidth]{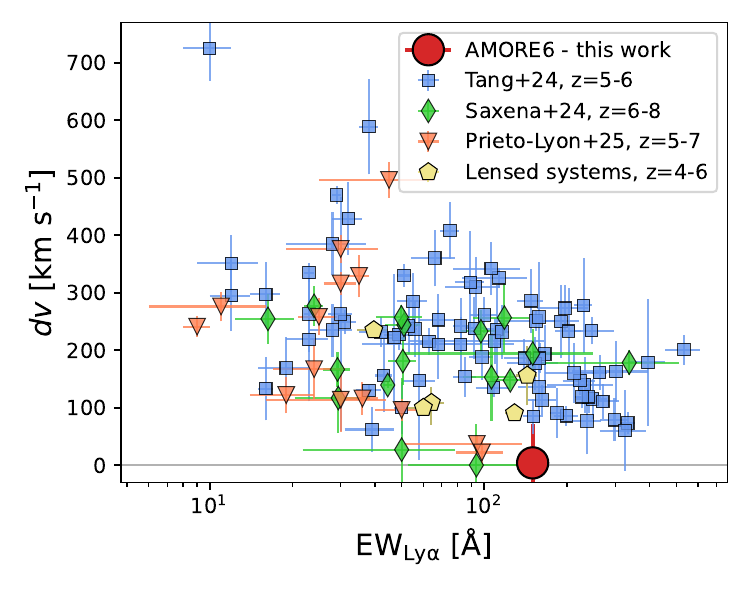}
\includegraphics[width=0.33\textwidth]{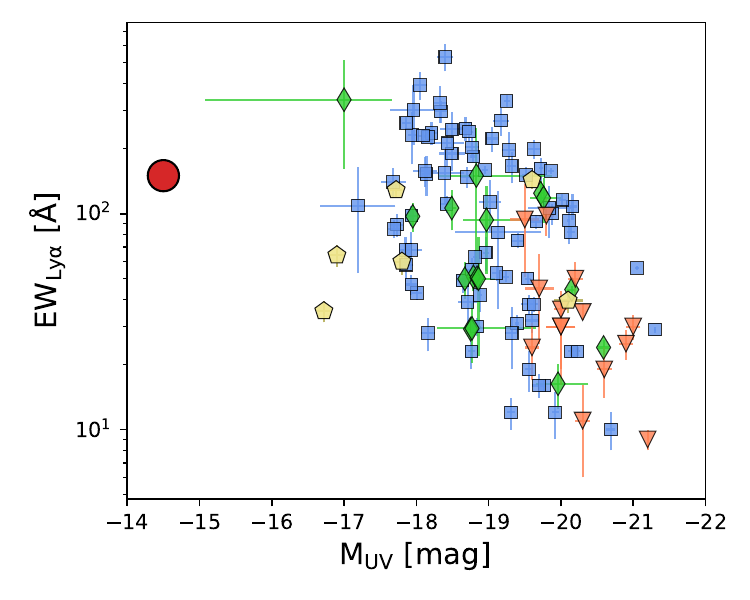}
\includegraphics[width=0.33\textwidth]{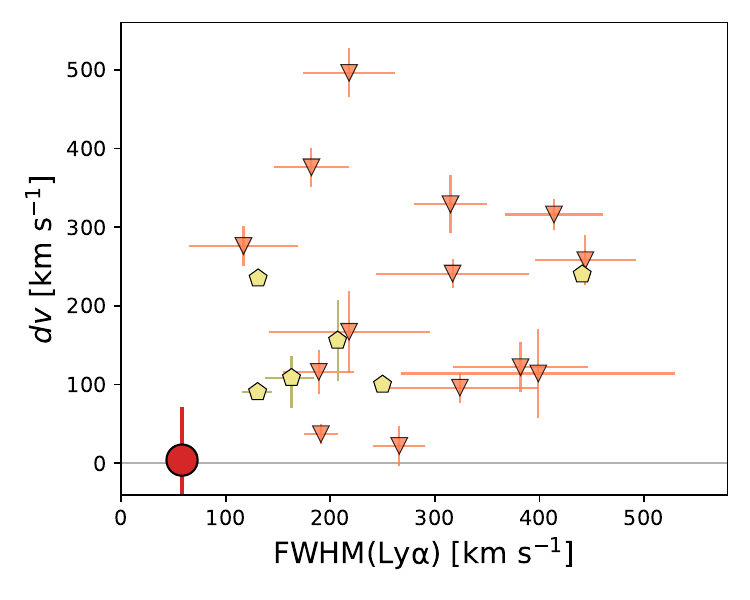}
\caption{Comparison of the \lya\ properties of AMORE6 with literature samples of galaxies at $\rm z>4$. {\it Left:} shift of the \lya\ peak relative to systemic ($dv$) versus rest-frame \lya\ EW; literature samples from \citet{Tang2024,Saxena2024,PrietoLyon2025_arXiv}; references for the lensed-galaxies sample in Table~\ref{tab:lensed}. {\it Center:} EW(\lya) versus $\rm M_{UV}$; AMORE6 is $4-5$ mag fainter than the comparison average. {\it Right:} $dv$ versus $\rm FWHM_{Ly\alpha}$ for cases with both measurements.}
\label{fig:literature}
\end{figure*}

Recent studies by \citet{Tang2024}, \citet{Saxena2024} and \citet{PrietoLyon2025_arXiv} put together a large sample of LAEs, including \lya\ line profiles, at similar redshifts to AMORE6. More specifically, \citet{PrietoLyon2025_arXiv} first provided a statistical sample of \lya\ FWHM at $\rm z\sim5$–$6$. The main \lya-related properties of these samples are shown in Fig.~\ref{fig:literature}. In addition various lensed LAEs in the same redshift range have been studied in the literature (though none intrinsically as faint as AMORE6); we collect their main properties and relative references in Table~\ref{tab:lensed}. Overall, while few galaxies have a \lya\ emission very close to the systemic redshift ($dv\sim0$ km/s), AMORE6 is the only case with a narrow \lya\ profile ($\rm FWHM<100~km/s$) robustly detected (right panel of Fig.~\ref{fig:literature}).
\begin{table*}[h!]
    \centering
    \renewcommand{\arraystretch}{1.35}
    \caption{\label{tab:lensed}\lya\ properties of lensed LAEs at $\rm z>4$ from the literature.}
    \begin{tabular}{l|lllll}
    ID & $\rm z_{sys}$ & $\rm M_{UV}$ & $\rm EW_{Ly\alpha}$ & $dv$ & FWHM \\
    \hline
    MACS0940    & 4.0338$^{(a)}$ & $-19.8^{(b)}$ & $-$ & $240\pm7^{(a)}$ & $441\pm8^{(a)}$$^\dagger$ \\
    RCS0224$^{(c)}$         & 4.8737   & $-19.6$ & $143.3\pm12.9$ & $156\pm52$ & $207\pm6$$^\dagger$ \\
    MS1358$^{(d)}$      & 4.9296   & $-19.0\pm0.1$ & $-$ & $200\pm100$ & $-$ \\
    CA8$^{(e)}$         & 6.064   & $-16.9\pm0.1$ & $64.2\pm7.4$ & $109^{+28}_{-38}$ & $163^{+22}_{-25}$ \\
    RXCJ2248$^{(f)}$ & 6.1045   & $-20.1\pm0.2$ & $39.6\pm5.1$ & $235$ & $131$ \\
    CA4$^{(e)}$         & 6.1446 & $-17.7\pm0.1$  & $128.8\pm9.4$ & $91^{+3}_{-4}$ & $131^{+14}_{-15}$ \\
    D1T1                    & 6.1449$^{(g)}$ & $-17.8\pm0.1^{(g)}$  & $60\pm8^{(h)}$ & $100^{(g)}$ & $250^{(g)}$$^\dagger$ \\
    CA5$^{(e)}$         & 6.149 & $-16.7\pm0.1$ & $35.4\pm4.0$ & $-$ & $176^{+110}_{-20}$ \\
    \hline
    \end{tabular}
    \tablefoot{Uncertainties are not reported when they are not available in the original publication. $^\dagger$~not corrected for instrumental broadening.}
    \tablebib{$^{(a)}$~\citet{claeyssens19}; $^{(b)}$~\citet{Messa2024}; $^{(c)}$~all data from \citet{Witstok2021}; $^{(d)}$~all data from \citet{Swinbank2009}; $^{(e)}$~all data from Bolamperti et al. in prep; $^{(f)}$~all data from \citet{Mainali2017}; $^{(g)}$~\citet{Messa_D1T1_2024}; $^{(h)}$~\citet{Vanzella2019}.} 
\end{table*}

\section{Comparison to radiative transfer models}\label{sec:app:RT}
\begin{figure*}[h]
\includegraphics[width=0.49\textwidth]{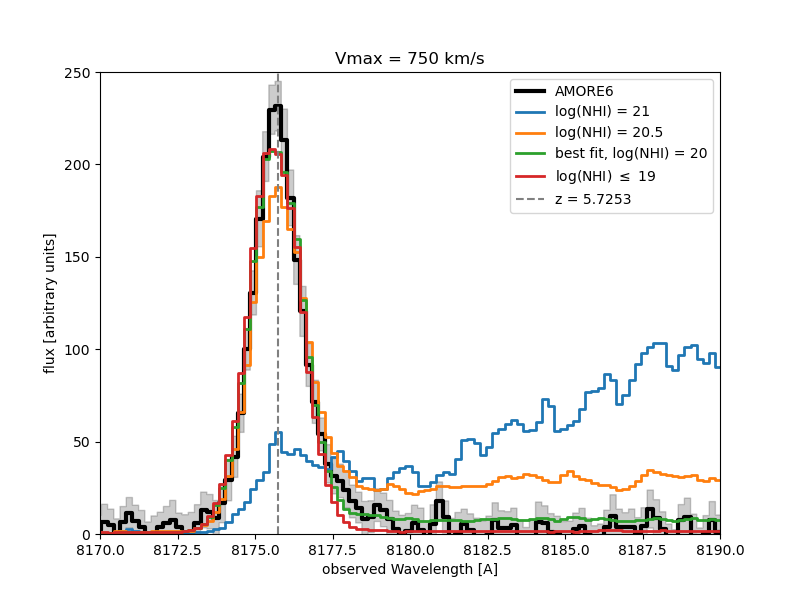}
\includegraphics[width=0.49\textwidth]{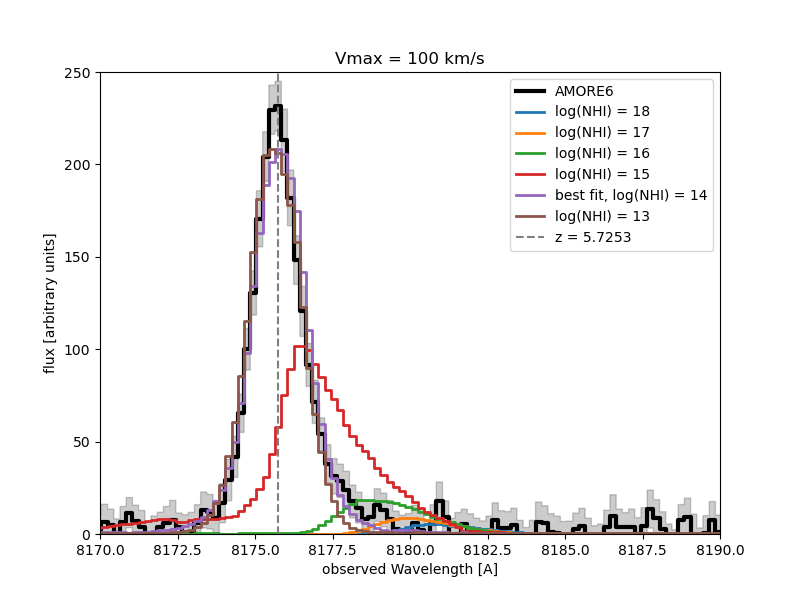}
\caption{Emergent \lya\ line profiles from a grid of idealized radiative-transfer models (see text). Left: high-velocity shell (V$_{\rm max} = 750$~km~s$^{-1}$); varying $\rm N_{HI}$ shows that 
$\rm log(N_{HI})\lesssim20$ reproduces the observed profile but requires an extreme outflow. Right: low-velocity case 
(V$_{\rm max} = 100$~km~s$^{-1}$); matching the data demands very low columns, 
$\rm log(N_{HI})\lesssim14$. Both panels assume the systemic redshift 
$z=5.7253$.}
\label{fig:RT}
\end{figure*}

We fit AMORE6 with the grid of idealized \lya\ radiative‐transfer models from \citet{Garel2024} accounting for the instrumental broadening and identified two families of acceptable solutions, both implying very low effective opacity: (i) very fast outflows, or (ii) extremely low H I columns. Figure~\ref{fig:RT} shows the best–fit models for each family, alongside variants in which all parameters are fixed to the best–fit values except $\rm N_{HI}$ (to illustrate its dominant impact on the \lya\ profile).

In the high-velocity family, the preferred solution adopts $\rm N_{HI}=10^{20}~cm^{-2}$ to reproduce the modest red wing; however faint residuals of this wing (almost consistent with the observational noise level) extend up to redder wavelengths than observed. Models with lower $\rm N_{HI}$ are essentially indistinguishable and recover the intrinsic line shape because the effective opacity is negligible, whereas higher $\rm N_{HI}$ suppress near–systemic photons and fail to match the data.
However, such an extreme outflow ($\rm V_{max}\approx750~km\,s^{-1}$) is probably unlikely for an \hii\ region/star cluster \citep[e.g.,][]{Turner2015,TenorioTagle2015}. Restricting the search to $\rm V_{max}\leq100~km\,s^{-1}$, good fits are obtained only with very low columns, with a best fit at $\rm N_{HI}\approx10^{14}~cm^{-2}$; larger densities inevitably shifts the \lya\ peak redward of systemic and degrades the fit. Under the hypothesis of low velocity of the expanding shell, this strongly indicates that the line–of–sight $\rm N_{HI}$ in AMORE6 is extremely low, consistent with a high LyC escape fraction.

A caveat is the IGM opacity at this redshift, which is not included in our forward models and could attenuate the blue side of the line. The observed profile’s near symmetry suggests this effect is modest, but it does not alter the conclusion that AMORE6 exhibits very low effective neutral columns.

\end{appendix}

\end{document}